\documentclass[doublecol,figures]{epl2} 

\usepackage{amsmath}
\usepackage{amsfonts}
\usepackage{graphicx}
\usepackage{braket}

\newcommand{\sigmaV}{{{\mbox{\boldmath$\sigma$}}}}
\newcommand{\D}{\Delta}
\newcommand{\w}{\omega}

\DeclareMathSymbol{\C}{\mathalpha}{AMSb}{'103}
\DeclareMathSymbol{\N}{\mathalpha}{AMSb}{'116}
\DeclareMathSymbol{\R}{\mathalpha}{AMSb}{'122}

\title{Decoherence induced by an ordered environment}
\author{Juliana Restrepo$^{1,2}$, S. Camalet$^{2}$, R. Chitra$^{2,3}$}
\institute{$^1$ Laboratoire de Physique Th\'eorique de la Mati\`ere Condens\'ee, UMR 7600, Universit\'e Pierre et Marie Curie, 4 place Jussieu, 75252 Paris Cedex 05, France. \\$^2$  
Grupo de Sistemas Complejos, Universidad Antonio Nari\~no, Medellin, Colombia\\ $^3$ Theoretische Physik, ETH Zurich, 8093 Zurich, Switzerland}

\abstract{ This Letter deals with the time evolution of a qubit  weakly coupled to a  reservoir which has a symmetry broken state
with long range order at finite temperatures.  In particular,  we model the ordered reservoir by a  standard BCS superconductor with
s-wave pairing.  We  study the reduced density matrix of a qubit  using  both the time-convolutionless  and Nakajima-Zwanzig
approximations. We study different kinds of couplings between the qubit and the superconducting bath.  We find that
 ordering in 
the superconducting  bath generically leads to 
an unfavorable non-Markovian faster-than-exponential decay of the qubit 
coherence. On the other hand,   a coupling of the qubit  to 
the non-ordered sector of the bath can result in a Markovian decoherence of the qubit  with a  drastic reduction of the  decoherence
rate.  Since these behaviors are endemic to 
the ordered phase, qubits can serve as useful probes of 
continuous phase transitions in their environment.  We also briefly discuss the   validity of our main result, faster than exponential 
decay of the qubit coherences,  for a  qubit coupled to  a generic ordered bath with a spontaneously broken continuous symmetry
at finite temperatures.  } 

\begin{document}

\maketitle

\section{Introduction}
The past decade has seen tremendous activity devoted to developing 
experimentally viable qubits for quantum computing. These two-level 
systems are realized, for example, by directly using the charge or spin 
degrees of freedom of electrons and quantum dots \cite{DiVincenzo1998} 
or more complex entities like flux qubits 
\cite{MSS,Johnson2011} and Cooper boxes \cite{Bouchiat1998}.
The utility of all these qubits for quantum computation  is strongly limited by 
the influence of their environment which tends to destroy their quantum 
coherence. Consequently, a lot of recent  theoretical  studies have
focused on ways and means of  increasing the coherence time scales 
\cite{Coish2008,Camalet2007a,Camalet2007b,Restrepo2011,
Milman2007,ioffe2002}. Another viewpoint  consists of employing such 
` ancillary qubits' as probes of the environment with which they interact, as typically done in NMR spectroscopy.  An  example  is  a spin-glass bath where,    the  reduced dynamics of the qubit was  shown to be directly sensitive to the spin glass order parameter\cite{Winograd2010}. This is especially useful in the field of   quantum optics, where given the difficulty  of standard thermodynamic measurements,
 such small probes can be used to  explore   phase transitions in Dicke like
models\cite{Brazil} and  also to 
  investigate both equilibrium and 
non-equilibrium properties 
of cold atoms\cite{Lunde2010}. 

The first theoretical studies of decoherence 
considered environments consisting
of independent harmonic oscillators \cite{LCDFGZ} or spins \cite{SH,FM}. 
More recently, complex baths have been studied. In particular, intrabath 
interactions have been taken into account
 \cite{Camalet2007b,Paganelli2002,Yuan2005, Rossini}, 
raising the question of the possible influence of thermodynamic phases and 
transitions on the decoherence of the qubit. 
It has been shown that the mitigating impact of intrabath interactions seen 
in many cases \cite{Paganelli2002,Yuan2005} breaks down when the bath 
is in the vicinity of a  phase transition\cite{Camalet2007a}.

Numerous studies \cite{Rossini, Sun2007, Cuc2010}  have addressed the question of  what happens when the bath is in the vicinity of a quantum phase transition  in the case of one dimensional  spin baths and have found enhanced decoherence near the critical point.  Note however, that the time evolution in these zero temperature systems is generically non-Markovian and due to the
one dimensional nature of the baths considered, no true long range order exists. 
 For higher dimensional baths at finite temperatures,  it was shown that the  Markovian decoherence rate diverges on the disordered side  as one approaches the second order transition temperature\cite{Camalet2007a} signaling the non-Markovian
time evolution in the ordered phase.
 Refs.\cite{Paganelli2002, Wang2008}  argued that once in the ordered phase, 
  symmetry breaking in the reservoir helps reduce decoherence, while 
 Ref.\cite{Yuan2005} found a strong Gaussian decay of quantum coherence.
 However,  these works suffer from different drawbacks ranging from 
 a complete neglect of low energy modes in Ref.\cite{Paganelli2002} to 
 obtaining an order parameter independent behavior for the
time evolution of the qubit in Ref.\cite{Yuan2005}.

In this paper, we revisit the problem  of ordered baths  at finite temperatures to have a clearer understanding 
of their effect   on qubits. We consider 
a superconducting bath at finite temperature in three dimensions  described by the Bardeen Cooper Schrieffer 
(BCS) theory. 
 Though the BCS Hamiltonian  does not  capture 
the fluctuations in the disordered phase, it provides a very good description of the ordered 
phase.  In line with Ref.\cite{Camalet2007a},  where it was shown that the 
impact of the ordering on the qubit depends crucially on
the  relation between the qubit-bath interaction and the order parameter,  we study 
different kinds of interactions between the qubit and the bath.
We shall show below that the ordered phase, characterized by a  spontaneously broken continuous symmetry,  is synonymous
with a non-Markovian time evolution of the qubit density matrix with interesting anomalous features.  Moreover,
the ordered  superconducting phase has  a rich variety 
of behaviors not seen in the disordered phase, including
faster-than-exponential decay of the coherence. 
The latter makes it unfavorable from the point 
of view of quantum computing.  But, there are interesting exceptional 
qubit states which decohere slower when the bath orders.
This sensitivity of the qubit to the order in the bath, makes it a good 
probe of the transition in the bath.

\section{Model}
The combined system of the qubit  and the superconducting bath is described by the Hamiltonian
\begin{eqnarray}
H&=&\sigmaV_q \cdot {\bf V} + H_B ,
 \label{H}
\end{eqnarray}
where $\sigmaV_q$  is the vector Pauli operator for the qubit, whose components are the usual 
2$\times$2 Pauli matrices, ${\bf V}$ is some bath vector operator that will be specified later, 
and $H_{B}$ is the conventional BCS Hamiltonian 
$H_{B}=\sum_{k\epsilon} E_k \alpha_{k\epsilon}^\dagger\alpha_{k\epsilon}$ 
\cite{Bardeen1957}. The Bogoliubov operators $\alpha_{k\epsilon}$ are related 
to the electron annihilation and creation operators by
\begin{equation}\label{A_Label}
 \begin{split}
\alpha^\dagger_{k\epsilon}=u_k c^\dagger_{k\epsilon} + v_k c_{-k-\epsilon}  \\
\alpha_{-k\epsilon}=u_k c_{-k\epsilon} - v_k c^\dagger_{k-\epsilon} ,
\end{split}
\end{equation}
where $c^\dagger_{k\epsilon}$ creates an electron with momentum $k$ and spin 
$\epsilon=\uparrow,\downarrow$. The BCS dispersion relation is 
$E_k= sgn(e_k)\sqrt{e_k^2 + \Delta^2}$, where $e_k$ is the underlying 
electronic dispersion and $\Delta$ is the superconducting 
gap. The coefficients in \eqref{A_Label} obey $(u,v)_k^2 = (1\pm e_k/E_k)/2$. 
We set $\hbar=k_B=1$ in the rest of the paper.The superconducting order parameter 
$\Delta$ at temperature $T$ is self-consistently determined by
\begin{equation}
g N \int_0^{\omega_D} 
de  \frac{\tanh(\sqrt{e^2 + \Delta^2}/2T)}{\sqrt{e^2 + \Delta^2}} = 1
\end{equation}
where $g$ is the strength of the phonon-mediated electron-electron interaction, 
$\w_D$ is the Debye frequency 
and $N$ is the electronic density of states at the Fermi surface.
As is well known, this equation determines a critical temperature $T_c$ which separates 
a high-temperature metallic phase where $\Delta=0$ and a low-temperature phase where $\Delta$ 
increases monotonically as $T$  decreases. 

We assume that, at time $t=0$, 
the qubit and the bath are uncorrelated and that the bath is in thermal equilibrium at temperature $T$. 
The initial state of the combined system is thus $\Omega=\rho(0)\otimes \rho_B$ 
where $\rho_B \propto \exp(-H_B/T)$ and $\rho(0)$ is any qubit density matrix. 
 The time evolution of the reduced density matrix 
of the qubit is given by 
\begin{equation}
  \rho (t) = \mathrm{Tr}_B[ {\rm e}^{-i Ht}\Omega {\rm e}^{i Ht}]
\end{equation}
    where $\mathrm{Tr}_B$  denotes the partial trace over 
the bath degrees of freedom. Since the superconducting bath has long range order, this implies that the typical correlation times of the bath are very long. This automatically precludes the use of  Markovian master equations, which relies on short bath correlation times, to study the time evolution of the reduced density matrix $\rho$.  Since, we anticipate non-Markovian behaviour in the present
problem,   in the
 limit of weak coupling between the qubit and the bath that we are interested in,  one can 
 use two main methods to calculate $\rho(t)$:  i)  the time-convolutionless (TCL) projection 
 operator technique and 
ii) the Nakajima-Zwanzig (NZ) approximation \cite{CDG,QDS}. Though both  methods can  deal with non-Markovian time evolution, TCL  gives local-in-time equations 
of motion  for $\rho(t)$ whereas the NZ approximation gives 
an integro-differential dynamical equation for $\rho(t)$.  The accuracy of these methods 
depends on the problem studied and it is  difficult to assert a priori which one is more 
appropriate \cite{BBP}. In this Letter, we focus on the asymptotic evolution predicted by 
the second-order TCL approximation, and briefly discuss the results obtained using 
the NZ technique at the end. To write the the master equation given by the TCL 
approximation to second order, it is convenient to first rewrite the Hamiltonian \eqref{H} as 
\begin{equation}
H=H_B+\sigmaV_q \cdot \langle {\bf V} \rangle+H_I
\end{equation}
\noindent 
 where 
$\langle \ldots \rangle=\mathrm{Tr}(\rho_B \ldots)$ and 
$H_I=\sigmaV_q \cdot ({\bf V}-\langle {\bf V} \rangle)$. With these notations, we obtain using the Born approximation
\begin{equation}\label{meq}
\partial_t \rho=-i[\sigmaV_q \cdot \langle {\bf V} \rangle ,\rho (t)] 
- \int_0^t d\tau \mathrm{Tr}_B \big[ H_I, [H_I(-\tau), \rho(t)\otimes\rho_B ] \big] ,
\end{equation}
 {where  the time dependent $H_I$ is given in the interaction picture by}
\begin{equation}
H_I(t)\equiv {\rm e}^{itH_B }H_I {\rm e}^{-itH_B}=\sigmaV_q \cdot ({\bf V}(t)-\langle {\bf V} \rangle)
\end{equation}

The TCL equation Eq.\eqref{meq} is our starting point for the   calculations which follow. We remark that, 
 replacing the upper limit of the integral  $t$ in \eqref{meq} by $+\infty$  leads  to the well-known 
Markovian master equation.  However, as we will show below,  
the time evolution of $\rho(t)$ can be non-Markovian in the ordered phase precluding the 
use of such Markovian master equations.  We note that  Markovian evolution means a  simple exponential   decay of the elements of the density matrix. To obtain the equivalent  equation  for the reduced density matrix within the NZ scheme, it suffices to replace $\rho(t)$ in the integral on the
right hand side of (\ref{meq}) by $\rho(t-\tau)$.  This results in a time non-local equation for the reduced density matrix.

\section{Kondo coupling}
We first consider a Kondo like coupling where the qubit couples 
to the electronic spin density at the origin ${\bf S}(0)$, i.e.,
\begin{equation}
V_\alpha=\lambda S_\alpha (0) \equiv \lambda 
\sum_{k,k',\epsilon,\epsilon'}c_{k\epsilon}^\dagger 
\sigma^\alpha_{\epsilon \epsilon'}  c_{k'\epsilon'}
\end{equation}
where $\alpha \in \{ x,y,z \}$, $\lambda$ is the coupling strength, and 
$\sigma^\alpha_{\epsilon \epsilon'}$ are the matrix elements of the Pauli matrix $\sigma^\alpha$. 
The Hamiltonian \eqref{H} is effectively the same as that of a magnetic impurity 
embedded in a superconductor. { Similar Kondo couplings were studied  experimentally   in multi-walled carbon nanotubes\cite{Schon} and  in spin half quantum dots coupled to 
two superconducting reservoirs in \cite{Lee}. } Contrary to the situation of a metallic bath where one really 
does not have a true weak coupling regime because of the dynamical Kondo effect, here we 
have a  weak coupling regime \cite{Chung1996}. 
The isotropy of the total Hamiltonian (\ref{H}) and the absence of any net moment in  the bath 
 lead to the following simplifications: since $\langle {\bf V} \rangle=0$, the first term in \eqref{meq} 
 vanishes and  all components of the effective spin-1/2 corresponding to the qubit, 
 i.e., $s_\alpha (t)= \mathrm{Tr} (\rho(t) \sigma_q^\alpha)$, satisfy the same equation of motion.
Consequently, the qubit evolution is characterized by a unique time function $M$ 
defined by 
\begin{equation}
s_\alpha (t)= M(t) s_\alpha (0)
\end{equation} 
 We see that  with a Kondo 
like coupling, both decoherence and relaxation exhibit the same time evolution. 
By writing the electron operators $c_{k\epsilon}$ in terms of the Bogoliubov quasi-particle 
operators \eqref{A_Label}, we find 
\begin{equation}
\ln  M(t) \simeq -4\lambda^2 \int_{-\infty}^{\infty} d\omega \dfrac{\sin (\omega t /2) ^2}{\omega^2} 
\Gamma^+(\w) , \label{eqTCL}
\end{equation}
 with
  \begin{equation}
  \label{gamma}
  \Gamma^\pm(\omega)=\left[S^\pm(\w)+S^\pm(-\w)\right]
  \end{equation} and
     and  the functions $S^\pm$  are given by
\begin{align}
S^{\pm}(\w)=&\int_{-\infty}^{\infty} de f^{\pm}(e,\w)\rho(e)\rho(e-\w)n(e) n(\w-e) ,
\label{sw}
\end{align}
where $f^{\pm}(e,\omega)\equiv 1\pm  {\D^2}/{e(e-\w)}$, 
$\rho(e)= \vert e \vert (e^2 -\Delta^2)^{-1/2}$ is the superconducting density of states, 
and  {$n(e)= 1/(\exp{( e/T)} +1)$} is the Fermi function. We note that $S^+(\w)$ and $S^-(\w)$  are the dynamical spin  and charge 
structure factors of
the superconducting bath. 

\begin{figure}
\centering
\includegraphics[width=7cm]{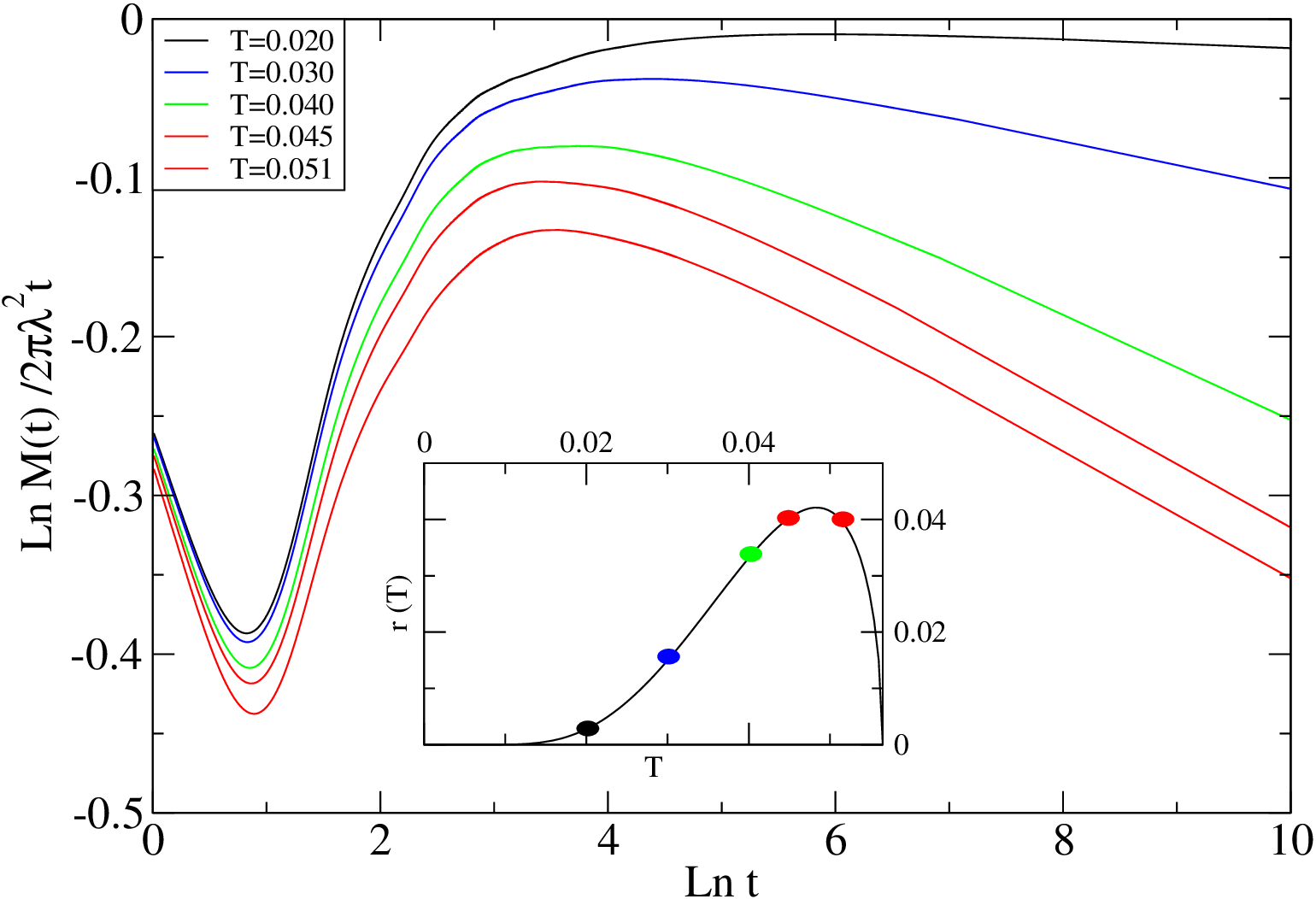}
\caption{ {\footnotesize $\ln M(t)/t$ as a function of $\ln t$ for $T=0.02$, $0.03$, $0.04$, 
$0.045$ and $0.051$ in units of $\w_D$. The inset shows the coefficient $r$ (in units of $\omega_D$) as a 
function of $T$. Here $gN=0.33$, $T_c\simeq0.056\w_D$ and $\Delta(0)\simeq0.1\w_D$.}}
\label{figura1}
\end{figure}

For $T > T_c$, the bath is a simple metal ($f^\pm(e)=1$) and we obtain the usual 
asymptotic Markovian decay $\ln M(t) = -\gamma t$  with a decay rate 
$\gamma=4\pi\lambda^2S^+(0)$ which increases and saturates to a density of states 
dependent value at very high temperatures \cite{Restrepo2011,Camalet2007b}. 
As one approaches the transition temperature $T \to T_c^+$, we expect the growing fluctuations 
to result in a divergent rate $\gamma$ at $T_c$ \cite{Camalet2007a} though this is 
not captured by the mean field  BCS theory used here.  We now analyze the asymptotic 
qubit evolution in the ordered phase  $0\leq T<T_c$. 
At $T=0$, we find that  $S^+(\omega) = 0 $ for all $\omega > -2\Delta$, 
where $2 \Delta$ is the gap to two particle excitations. 
{ {This gap leads to an incomplete decoherence of the qubit  where the function   $M(t)$  approaches a constant as a power law in the asymptotic limit    $t \to \infty$ 
just as seen in the case of an insulating bath \cite{Restrepo2011}.   The resulting   asymptotic density matrix $\rho$ does not lose
the memory of the initial conditions since $M(t) \neq 0$ implying that  the qubit is not in a simple statistical mixture  with equal probabilities of spin up and spin down states.}}
At  temperature $0< T<T_c$,  due to the divergence present in  the superconducting 
density of states, 
$S^+$ is infra-red divergent: $S^+(\w)\simeq -r(T) \ln|\w/T|$ as $\w \rightarrow 0$.
This divergence stems from the existence of Goldstone modes in the ordered phase.
The pre-factor $r(T)=(\Delta/2)\cosh^{-2}(\Delta/2T)$, shown in the inset of Fig.\ref{figura1}, 
is a non-monotonic function of temperature which vanishes at $T=0$ and $T=T_c$.
This infrared divergence results in 
\begin{equation}
{\ln M(t)} \simeq -{2\pi\lambda^2} r(T)  t \ln t
\label{tcl}
\end{equation}
for times $t \gg  t_a\equiv \mathrm{Max} (1/T,1/\D)$.
We  see that, contrary to naive expectations,  the ordered bath  leads to a novel 
faster-than-exponential loss of coherence of the qubit.  This can be attributed to the fact 
that the qubit couples to the spin fluctuations and hence order parameter fluctuations 
via the singlet Cooper pairs. 
Moreover, we see an interesting  reentrance in the asymptotic 
regime because the coefficient $r(T)$ which dictates the asymptotic decoherence is 
the same for two different temperatures (cf Fig. \ref{figura1}).  

\begin{figure}
\centering
\includegraphics[width=7cm]{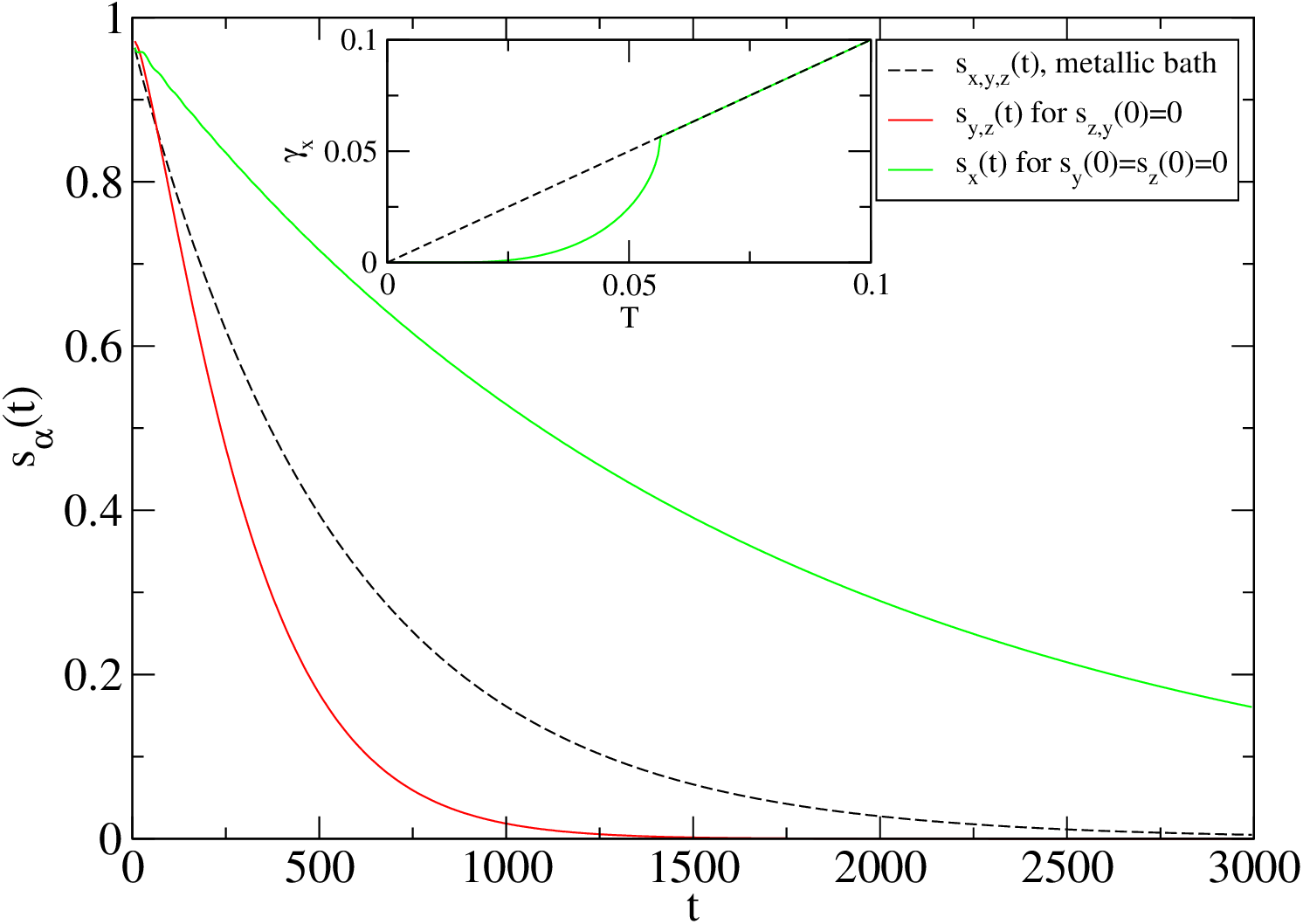}
\caption{ {\footnotesize $s_\alpha(t)$ as a function of $t$ for
the order coupling to a superconducting bath at $T=0.045\w_D$ for the initial 
conditions indicated in the legend. For the components $s_y$ and $s_z$, 
only the envelope is plotted. The inset shows $\gamma_x$ ( in units of $4\pi \lambda^2$) as a function of $T$  (in units of $\omega_D$) for metallic 
and superconducting baths. Here $gN=0.33$ and $\lambda=0.05\omega_D$.}} 
\label{figura2}
\end{figure} 

\section{Order coupling}
To further understand the physical origin of the ultra-fast decoherence seen above for a Kondo coupled qubit, we
  now consider  a direct coupling of the qubit  to the order parameter of the bath  of the form
\begin{equation}\label{orderint}
\lambda [\sigma_q^- c^\dag_{0\uparrow}c^\dag_{0\downarrow} + \sigma_q^+ c_{0\downarrow}c_{0\uparrow}]
\end{equation}
where $\sigma_q^\pm$ are the qubit spin raising and lowering operators and $c^\dag_{0\uparrow}c^\dag_{0\downarrow}$
is the superconducting order parameter at the origin.
 This coupling can be rewritten{ { in momentum space }} and  in terms of the operators
$V$ such that $V_z=0$ and 
\begin{eqnarray}
V_x &=&\lambda\sum_{k,k'}(c_{k\uparrow}^\dagger c^\dagger_{k' \downarrow} + c_{k' \downarrow}c_{k\uparrow}) \nonumber \\
V_y&=&i\lambda\sum_{k,k'}(c_{k\uparrow}^\dagger c^\dagger_{k' \downarrow} - c_{k' \downarrow}c_{k\uparrow} ) 
\end{eqnarray}
These operators are directly related to the superconducting order parameter  operator at the origin, $O=\sum_{k,k'}c_{k\uparrow}^\dagger c^\dagger_{k' \downarrow}$. {{Such couplings can be realized in  qubits made from 
Cooper pair boxes which are capacitively tunnel coupled to a superconducting reservoir \cite{Nakamura1999,Bouchiat1998}. In this
case, the interaction Hamiltonian describes the tunneling of a Cooper pair from the box into the reservoir and vice versa.
In general, depending on the details of the single electron tunneling coefficients, the interaction term in Eq.\eqref{orderint} would be modified by  suitable form factors. Multi-qubit models with similar interactions with a superconducting reservoir have also been
proposed  to observe quantum optics phenomena like superradiance in mesoscopic systems \cite{Rodriguez}. In our view,  the case  of a single qubit   still remains to be   fully understood.}} Consequently,  for  the bath in thermal equilibrium,  as opposed to the Kondo coupling case, here we have $\langle V_y \rangle = \langle V_z \rangle =0$
and $\langle V_x \rangle \neq 0$.
In this case, we obtain, from \eqref{meq}, the following set of coupled differential equations 
for the components $s_\alpha$:
\begin{equation}
\partial_t \left(\begin{array}{c}
s_x\\
s_y\\
s_z \end{array}\right)=
\left( \begin{array}{cccc}
\Sigma_-(t) & 0 & 0 \\
0 & \Sigma_+(t) & h \\
 0 & -h & \Sigma_-(t)+\Sigma_+(t)\end{array}\right)
\left( \begin{array}{c}
s_x\\
s_y\\
s_z \end{array} \right)
\label{ecuacionM}
\end{equation}
The  first order term $h\equiv -2\braket{V_x}= 2 \lambda\Delta/g N$ and the  functions $\Sigma_\pm$ are given by
\begin{equation}
\Sigma_\pm(t)= -4\lambda^2 \int_{-\infty}^{\infty} d\w\frac{\sin(\w t)}{\w}\Gamma^\pm(\w)
\end{equation}  
with the $\Gamma^\pm$ defined earlier in (\ref{sw}).
 Unlike the Kondo case previously studied, 
here the equations for the components $y$ and $z$ 
are coupled in the ordered phase where $\Delta \ne 0$. 
A closer look shows that $\Sigma_-$ and hence $s_x$ 
are related to the  dynamic charge correlation function  which is non-singular in the superconducting phase and 
$\Sigma_+$ to the spin correlation function which shows singular behavior in the superconducting phase.
 As will be discussed  below, this  leads to a complex time evolution for the different components of the qubit density
 matrix.

For $T > T_c$, all the components  $s_\alpha$ are uncoupled  and decay asymptotically as 
$\ln s_\alpha  =  -\gamma_\alpha t$  with  the 
rates $\gamma_x= \gamma_y = 8\pi\lambda^2 S^-(0)$ and 
$\gamma_z= 2 \gamma_x$. The existence of two different rates is a direct 
consequence of the spin anisotropy of the order coupling whereas in the spin 
isotropic Kondo case, all rates are the same. In the ordered phase, at $T=0$, 
solving Eq.\ref{ecuacionM}, we find that  the presence of a gap, both in $\Sigma_-$ 
and $\Sigma_+$, leads to an incomplete decay of the central spin coherences.  For $T \neq 0$,  since
$S^-$ is always regular  and finite at low frequencies  this results in a Markovian decay 
$\ln {s_x}  \simeq -\gamma_x t $ for times $t\gg t_a$. The full temperature 
dependence of the rate $\gamma_x$ is shown in the inset of Fig. \ref{figura2}. 
Its behavior for $T\to 0$ is $\gamma_x \propto T e^{-\Delta/T}$. 
This is an important result 
of our work because it shows that in the ordered phase, the Markovian rate 
for the component $s_x$ is strongly suppressed compared to a simple 
metallic bath.  This is very similar to the
  relaxation induced by  a coupling to the charge fluctuations, studied in the context 
  of NMR by Fulde and Black \cite{Fulde1979}, which is described by 
$\sigmaV_q \cdot {\bf V}\propto \sigma_z {\hat n}$ 
  where ${\hat n}$ is the number of electrons at the origin. 
For $T\geq T_c$, the rate $\gamma_x$ coincides with the rate of the equivalent metallic 
bath with $\Delta=0$ and saturates to a finite DOS-dependent value proportional 
to $\int dE \rho(E)^2$ as expected. 
 If the qubit's  entire evolution was determined by  this component, then the bath is 
 effectively a semiconductor with a temperature dependent gap \cite{Restrepo2011}.   
 On the other hand, we find that the components $s_y$ and $s_z$ exhibit non-Markovian behavior. 
In the limit of weak coupling between the qubit and the bath, Eq.\eqref{ecuacionM} can
 be solved for $s_y$ and $s_z$ \cite{Rscheisen2006}. { { The full  solution  is given by  $(s_y(t),s_z(t))=(s_y(0),s_z(0))\exp(A)$ 
 where the  $2\times 2$ matrix  $A$ is typically an infinite series in the coefficient $\lambda$. However, since we are interested in the weak coupling limit,  it suffices to retain terms only  upto second order in the coupling $\lambda$ in the matrix  $A$. This is equivalent to the TCL result for the Kondo coupling seen in the previous section (cf. Eq.\eqref{eqTCL}).  Doing the algebra, we find, }} 
\begin{align}
s_y\simeq &\,e^{a_1}\Big[s_y(0)\left(a_0\mathrm{sinc}\, \Theta +\cos \Theta \right)+s_z(0)ht \, \mathrm{sinc}\, \Theta\Big]\label{myt}\\
s_z\simeq &\,e^{a_1}\Big[-s_y(0) ht \, \mathrm{sinc}\, \Theta + s_z(0) \left(\cos{\Theta}-a_0\mathrm{sinc}\, \Theta\right) \Big]\label{mzt}\nonumber
\end{align}
 {where the time dependent functions $a_0$ and $a_1$ are given by}
\begin{eqnarray}
 a_\mu(t) &= & \int_0^t{ dt'[\mu \Sigma_+(t')+(\mu-1/2) \Sigma_-(t')]}  \\
\Theta(t) &=& [(ht)^2-a_0(t)^2]^{1/2}\simeq ht
\end{eqnarray}
\noindent
 and $\mathrm{sinc}\, \Theta=\Theta^{-1}\sin \Theta$.  
Note that both $s_y$ and $s_z$ show oscillatory behavior in the ordered phase with 
a frequency proportional to the order parameter.  At $T=0$, $s_y$ and $s_z$ oscillate at 
a maximal frequency and  at finite temperatures $0< T<T_c$, 
 these oscillations are damped with the  envelope  $a_\mu(t)\propto - r(T) t\ln t$  for  
 $t\gg t_a$ and we recover the faster-than-exponential decay (\ref{tcl}) 
 with reentrant temperatures encountered for the Kondo coupling (cf Fig. \ref{figura2}).  
 The appearance of  oscillations of the qubit  can be used as a tool to demarcate the
  phase diagram of the superconductor.
 
Because of the very different asymptotic behaviors of the components $s_\alpha$ 
found above, the decay timescale of the qubit depends crucially on the initial conditions. 
Assume the qubit is initially prepared in an eigenstate of $\sigma_x$. 
Then the components $s_y(t)=s_z(t)=0$ are constant, and the asymptotic 
time evolution of the qubit, determined by $s_x(t)$, 
is non-oscillatory and Markovian with 
a highly reduced rate in the ordered phase. 
These pure states are thus relatively stable in the environment 
considered here. Moreover, a combination of good initial 
preparation and pulse sequences which repeatedly orient the qubit in the $x$ 
direction could efficiently reduce decoherence times \cite{Morgan2008,Lutchyn2008}.
For an initial qubit state such that $s_x(0)=0$, this component remains zero 
and the asymptotic behavior of the qubit consists of an oscillatory and 
faster-than-exponential decay. 
In this case, due to the coupling between $y$ and $z$ components, 
we have situations where even if one of the components 
$s_y$ or $s_z$ is initially zero, this component can grow with time and 
show oscillatory behavior. For generic initial states, 
since $s_y$ and $s_z$ decay much faster than $s_x$, the asymptotic 
time evolution of the qubit is Markovian. The qubit state first decoheres to 
a statistical mixture of the eigenstates of $\sigma_x$ and then relaxes 
exponentially into the maximally mixed state.

\begin{figure}
\centering
\includegraphics[width=7cm]{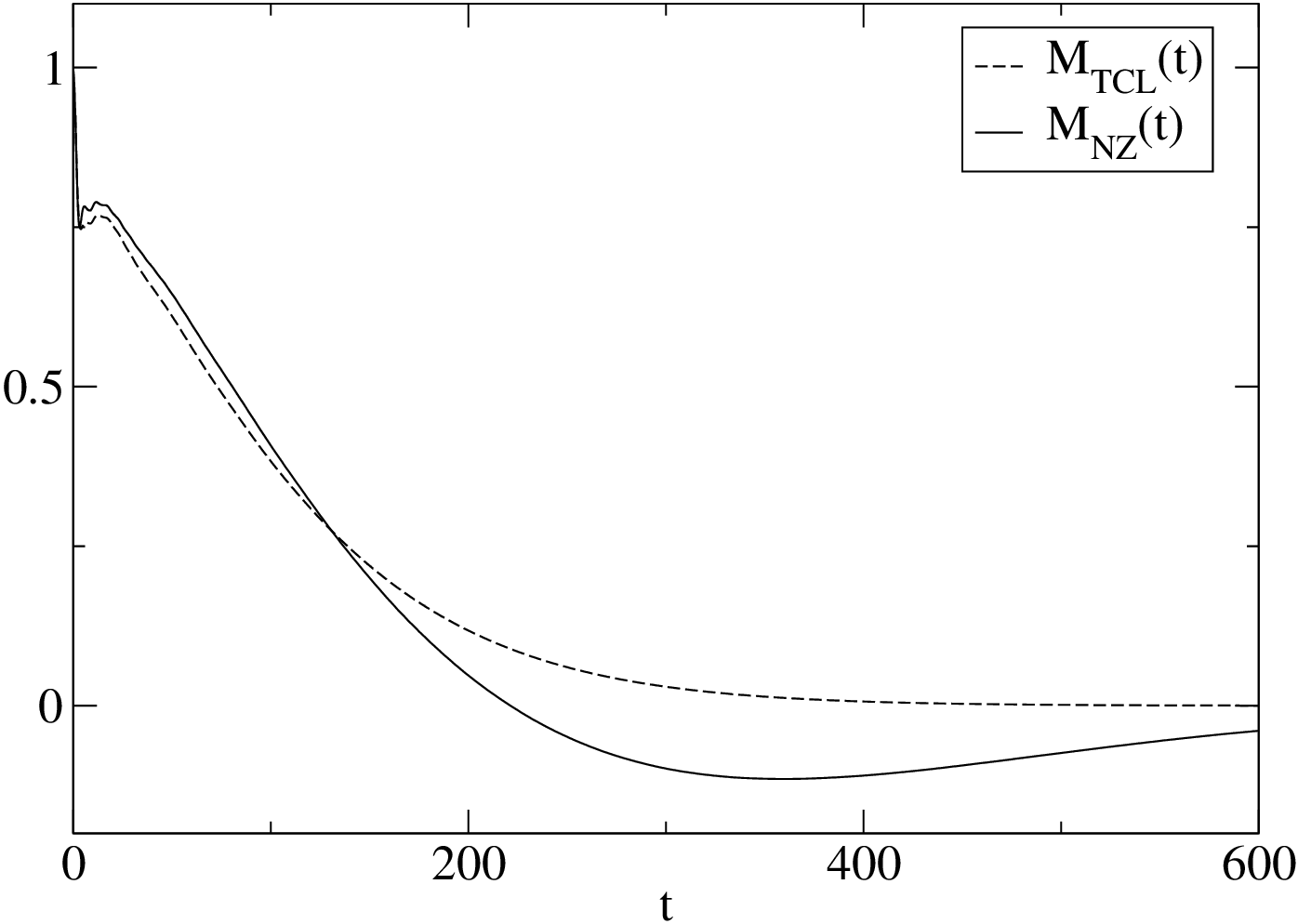}
\caption{\footnotesize $ M(t)$ as a function of $ t$ for the Kondo coupling case using  both TCL and NZ approximations.
The parameters are $T=0.03 \w_D$, $\lambda=0.07 \omega_D$, $gN=0.33$ and  $T_c\simeq0.056\w_D$ .}
\label{fignz}
\end{figure}

In both the Kondo and order coupling cases, we find interesting intermediate 
time behaviors but these will be discussed elsewhere.  We now 
briefly discuss the results obtained using the NZ approximation \cite{Restrepo2011}. 
To second order, the reduced density matrix satisfies (\ref{meq}) with $\rho_s(t)$ in the
integral in the second term replaced by $\rho_s(t-\tau)$.  Using the results for $S_\pm$
given in Eq.(\eqref{sw}), we find  that in the case of the  Kondo coupling, the asymptotic 
faster-than-exponential behavior seen earlier is replaced
 by a much slower non-Markovian behavior  
$
M_{nz}(t)\sim\int d\w\cos(\w t)/\ln|\w|  \sim -1/t\ln{t}$ for  times $t \gg t_{nz}$. 
We find  $t_{nz} \gg  t_a$  and an intermediate regime $t_a \ll t \ll t_{nz}$ 
characterized by a quantitatively faster  decay  than that predicted by the 
TCL approach, where the qubit becomes practically incoherent.
In the case of the order coupling, we find that 
the conclusions for the asymptotic  Markovian behavior seen for the component $s_x$  remain unchanged
whereas,
 the components $s_y$ and $s_z$   show the same functional form of decay given by  $M_{nz}$.
 We plot the full time evolution predicted by both approximation methods in Fig.\ref{fignz}.
 We can safely conclude that
anomalously fast decoherence seems to be a feature of both NZ and TCL 
methods in the ordered phase.  

\section{Conclusion}
To summarize, we have studied the influence of a true long-range ordered bath 
on the state of a qubit.  
For two different qubit-bath couplings  and generic initial conditions, we found a faster-than-exponential decoherence 
of the qubit in the ordered phase leading to 
the conclusion that ordered baths are often disastrous for qubits.  This  non-Markovian time
evolution essentially stems
from the contribution of the Goldstone modes to the bath correlation functions in the
symmetry broken phase.
{ {However, for  a direct coupling  of the qubit to the ordering operator, 
some particular pure states of the qubit  with initial values $\langle s_y\rangle = \langle s_z\rangle =0$ lead to pure decoherence with a highly reduced Markovian decoherence rate  when the bath orders,}} making such states potentially useful 
in experiments using echo sequences or other related techniques. 
This result constitutes another example of the important role played by the  nature of 
 the coupling to a bath that can order \cite{Camalet2007a}. 
For generic initial states of the qubit, we expect, if fluctuations in the disordered phase are properly taken into account, 
a divergence of the Markovian rate at the transition from the disordered side, 
followed by a faster-than-exponential loss of coherence in the ordered phase.
Exceptions are possible, as our study shows. 

We believe that this picture  of non-Markovian  time evolution in the ordered phase should be valid for all ordered baths 
where a continuous symmetry is spontaneously broken, 
provided the qubit couples in some way to the order parameter.   
In the problem studied here, the  faster than exponential non-Markovian behaviour in the ordered phase is
intricately linked to the existence of the soft Goldstone modes in the ordered phase
due the the spontaneous breaking of the continuous $U(1)$ symmetry.   Typically, the  non-Markovianity  could be
a  faster than exponential decay or a power law decay. Power-law decay  (seen  for example, in the spin boson problem at $T=0$) however, would
imply slow decoherence.  To obtain such power law decays for the case or ordered baths at finite temperatures would
require a $\Gamma^\pm(\omega)  \propto \vert \omega \vert$.  Given that $\Gamma^\pm$ is related to the dynamical structure
factors of the bath, $S^\pm(\w)$, fluctuation dissipation theorem tells us that  $S^\pm(\w=0) \neq 0$ at finite temperatures.
This effectively rules out any power law asymptotic behaviour   and any resulting non-Markovian behaviour has to be faster
than exponential.  The full quantitative time evolution will ofcourse depend on the particular physical problem studied. 
 An obvious question
is, whether one would obtain such enhanced non-Markovian  loss of coherence in
other ordered systems
where  the order arises  from a spontaneously broken discrete symmetry like $Z_N$  as in the latter one would not expect the formation of 
Goldstone modes.  It would be interesting to generalize our model to a non-degenerate qubit and 
to study the effect of such anomalous dissipation on the tunneling of 
the qubit \cite{LCDFGZ}. These question are beyond the scope of the present paper and are left for future work.


\begin{thebibliography}{23}

\expandafter\ifx\csname natexlab\endcsname\relax\def\natexlab#1{#1}\fi
\expandafter\ifx\csname bibnamefont\endcsname\relax
  \def\bibnamefont#1{#1}\fi
\expandafter\ifx\csname bibfnamefont\endcsname\relax
  \def\bibfnamefont#1{#1}\fi
\expandafter\ifx\csname citenamefont\endcsname\relax
  \def\citenamefont#1{#1}\fi
\expandafter\ifx\csname url\endcsname\relax
  \def\url#1{\texttt{#1}}\fi
\expandafter\ifx\csname urlprefix\endcsname\relax\def\urlprefix{URL }\fi
\providecommand{\bibinfo}[2]{#2}
\providecommand{\eprint}[2][]{\url{#2}}


\bibitem{DiVincenzo1998}
\Name{D. P. DIVINCENZO and D. LOSS}
\REVIEW{Superlattices and Microstructures}{\bf 23}{1998}{419}.

\bibitem{MSS}
\Name{Y. MAKHLIN, G. SCH\"ON and A. SHNIRMAN}
\REVIEW{Rev. Mod. Phys. }{\bf 73}{2001}{357}.

\bibitem{Johnson2011}
\Name{M.W. JOHNSON et al}
\REVIEW{Nature}{\bf 473}{2011}{194}.

\bibitem{Bouchiat1998}
\Name{V. BOUCHIAT et al}
\REVIEW{Physica Scripta}{\bf 76}{1998}{165}.

\bibitem{Coish2008}
\Name{W.A. COISH, J. FISCHER and D. LOSS}
\REVIEW{Phys. Rev. B}{\bf 77}{2008}{125329}.

\bibitem{Camalet2007a}
\Name{S. CAMALET  and R. CHITRA}
\REVIEW{Phys. Rev. Lett.}{\bf 99}{2007}{267202}.

\bibitem{Camalet2007b}
\Name{S. CAMALET  and R. CHITRA}
\REVIEW{Phys. Rev. B}{\bf 75}{2007}{094434}.

\bibitem{Restrepo2011}
\Name{J. RESTREPO, S. CAMALET, R. CHITRA  and E. DUPONT}
\REVIEW{Phys. Rev. B}{\bf 84}{2011}{245109}.

\bibitem{Milman2007}
\Name{P. MILMAN et al}
\REVIEW{Phys. Rev. Lett.}{\bf 99}{2007}{020503}.

\bibitem{ioffe2002}
\Name{G. IOFFE et al}
\REVIEW{Nature}{\bf 415}{2002}{503}.

\bibitem{Winograd2010}
\Name{E.A. WINOGRAD, M.J. ROZENBERG and R. CHITRA}
\REVIEW{Phys. Rev. B}{\bf 80}{2009}{214429}.

\bibitem{Brazil}
\Name{J.P. SANTOS, F.L. SEMIAO and K. FURUYA} 
\REVIEW{Phys. Rev. A}{\bf 82}{2010}{063801}.

\bibitem{Lunde2010}
\Name{A.M. LUNDE, S. E. NIGG and  M. BUTTIKER}
\REVIEW{Phys. Rev. B}{\bf 81}{2010}{041311}.


\bibitem{LCDFGZ} 
\Name{A.J. LEGGETT et al}
\REVIEW{Rev. Mod. Phys.}{\bf 59}{1987}{1}. 

\bibitem{SH} 
\Name{J. SHAO and P. H\"ANGGI}
\REVIEW{Phys. Rev. Lett.} { \bf 81}{1998}{5710}.

\bibitem{FM} 
\Name{K.M. FORSYTHE and N. MAKRI}
\REVIEW{Phys. Rev. B} {\bf 60}{1999}{972}. 

\bibitem{Paganelli2002} 
\Name{S. PAGANELLI, F. dePASQUALE and S.M. GIAMPAOLO}
\REVIEW{Phys. Rev. A.} {\bf 66}{2002}{052317}. 

\bibitem{Rossini}
\Name{D. ROSSINI et al}
\REVIEW{Phys. Rev. A}{ \bf 75}{2007}{ 032333}
\bibitem{Yuan2005} 
\Name{XIAO-ZHONG YUAN and KA-DI ZHU}
\REVIEW{Europhysics Letters.} {\bf 69}{2005}{868}. 

\bibitem{Sun2007}
\Name{Z. SUN, X. WANG and C.P. SUN }
\REVIEW{Phys. Rev. A}{ \bf 75}{2007} {062312}.

\bibitem{Cuc2010}
\Name{F. M. CUCCHIETTI et al}
\REVIEW{Phys. Rev. Lett.}{\bf  105}{2010}{ 240406}.


\bibitem{Wang2008} 
\Name{X.X. YI, L.C. WANG and H.T. CUI}
\REVIEW{Physics Letters A} {\bf 372}{2008}{1387}. 

\bibitem{Bardeen1957} 
\Name{J. BARDEEN, L.N. COOPER and J.R. SCHRIEFFER}
\REVIEW{Europhysics Letters.} {\bf 108}{1957}{1175}. 

\bibitem{CDG}
\Name{C. COHEN-TANNOUDJI, J. DUPONT-ROC and G. GRYNBERG}
\Book{Processus d'interaction entre photons et atomes} 
\Publ{CNRS Editions, Paris} 
\Year{1988}.

\bibitem{QDS} 
\Name{WEISS U.}
\Book{Quantum dissipative systems} 
\Publ{World Scientific, Singapore}
\Year{1993}.

\bibitem{BBP} 
\Name{H.-P. BREUER, D. BURGRATH and F. PETRUCCIONE}
\REVIEW{Phys. Rev. B}{ \bf 70}{2004}{045323}.

\bibitem{Chung1996} 
\Name{W. CHUNG and M. JARRELL}
\REVIEW{Phys. Rev. Lett.}{\bf 77}{1996}{3621}.


 {\bibitem{Schon}
\Name{M. R. BUITELAAR, T. NUSSBAUMER and C. SCHOENENBERGER}
\REVIEW{Phys. Rev. Lett }{\bf 89}{2002}{ 256801}.

\bibitem{Lee}
\Name{EDUARDO J. H. LEE  et al }
\REVIEW{Phys. Rev. Lett}{\bf 109}{2012}{186802}.}

\bibitem{Nakamura1999} 
\Name{Y. NAKAMURA, Yu. A. PASHKIN and J.S. TSAI}
\REVIEW{Nature}{\bf 398}{1999}{786}.


 {\bibitem{Rodriguez}
\Name{D A RODROGUES, B L GYORFFY and T P SPILLER}
\REVIEW{J. Phys.: Condens. Matter}{\bf 16} {2004} {4477}.}


\bibitem{Fulde1979} 
\Name{J.L. BLACK and P. FULDE}
\REVIEW{Phys. Rev. Lett}{\bf 43}{1979}{453}.

\bibitem{Rscheisen2006} 
\Name{C. RSCHEISEN, W. BALSER and F. STEINER}
\REVIEW{Ulmer Seminare ber Funktional-analysis und Differentialgleichungen}{11}{2006}{53}.

\bibitem{Morgan2008} 
\Name{S.W. MORGAN, B.V. FINE and B. SAAM}
\REVIEW{Phys. Rev. Lett}{\bf 101}{2008}{067601}.

\bibitem{Lutchyn2008} 
\Name{R. LUTCHYN et al}
\REVIEW{Physica Scripta}{78}{2008}{024508}.



\end{thebibliography}
\end{document}